\documentclass[3p,times,twocolumn]{elsarticle}
\usepackage{ecrc}

\volume{00}

\firstpage{1}

\journalname{Nuclear Physics B Proceedings Supplement}

\runauth{M.Pepe Altarelli}

\jid{nuphbp}

\jnltitlelogo{Nuclear Physics B Proceedings Supplement}

\usepackage{amssymb}
\usepackage{amsmath}
\usepackage{hyperref}     
\usepackage{graphicx}
\usepackage{epstopdf}

\newcommand{\PT}{p_{\mathrm{T}}}
\newcommand{\bb}{$b\overline{b}$}
\newcommand{\cc}{$c\overline{c}$}
\newcommand{\lum}{\mathcal{L}}
\newcommand{\cms}{\mbox{cm$^{-2}$ s$^{-1}$}}
\newcommand{\Jpsi}{\ensuremath{J\mskip -2.5mu/\mskip -1mu\psi\mskip 1mu}}
\newcommand{\BBar}{$B^0_s$-${\kern 0.18em\overline{\kern -0.18em B^0_s}}$}
\newcommand{\BJPSIPHI}{$B^0_s \rightarrow \Jpsi \phi$}

\newcommand{\BJPSIKK}{$B^0_s  \rightarrow \Jpsi  K^+ K^-$}
\newcommand{\BS}{$B_s$}
\newcommand{\BC}{$B_c$}
\newcommand{\BbarS}{$\overline{B}_s$}

\begin{document}
\begin{frontmatter}




\title{$CP$ violation in charm and beauty decays at LHCb}


\author{M. Pepe Altarelli\fnref{label2}}
\ead{monica.pepe.altarelli@cern.ch}
\address{CERN, Geneva, Switzerland}

\fntext[label2]{On behalf of the LHCb collaboration}

\begin{abstract}
LHCb is a dedicated heavy flavour physics precision experiment at the LHC searching for New Physics (NP) beyond the Standard Model (SM) through the study of very rare decays of beauty and charm-flavoured hadrons and precision measurements of $CP$-violating observables. In this review I will present a selection of recent precision measurements of $CP$-violating observables in the decays  of beauty and charm-flavoured hadrons. These measurements are  based on an integrated luminosity of up to $1.0~\rm{fb}^{-1}$ collected by LHCb in 2011. 

\end{abstract}

\begin{keyword}
LHCb \sep Flavour physics\sep CP violation

\vspace*{0.5cm}
\line(1,0){466}
\begin{center}
\hfil{LHCb-PROC-2013-002\,; CERN-LHCb-PROC-2013-002\phantom{xxxxxxxxx}}\\ 
\vspace*{0.3cm}
\hfil{\small Fourth Workshop on Theory, Phenomenology and Experiments in Flavour Physics\phantom{xxxxxxxxx}}\\
\hfil{\small 11 - 13 June 2012, Anacapri, Italy\phantom{xxxxxxxxxxxxxxxxxx}}\\
\end{center}

\end{keyword}

\end{frontmatter}


\let\thefootnote\relax\footnotetext{\copyright~CERN on behalf of the LHCb collaboration, license \href{http://creativecommons.org/licenses/by/3.0/}{CC-BY-3.0}.}

\section{Introduction}
\label{intro}
The LHCb detector has been taking data with high efficiency during the last three years of operation at the LHC, producing a wealth of exciting physics results, which have made an impact on the flavour physics landscape and proved the concept of a dedicated heavy flavour spectrometer in the forward region at a hadron collider. 

The LHC is the world's most intense source of $b$ hadrons. The  \bb\ cross-section in proton-proton collisions at  $\sqrt{s}=7$~TeV is measured to be $\sim$$300~ \mu$b, implying that more than $10^{11}$ \bb\ pairs were produced in LHCb  in 2011, when an integrated luminosity of $1.0\,\rm{fb}^{-1}$ was collected.  The \cc\ cross-section is about 20 times larger than the \bb\ cross-section,  giving LHCb great potential in charm physics studies. As in the case of the Tevatron, a complete spectrum of $b$ hadrons is available, including \BS, \BC\ mesons and $b$ baryons, such as $\Lambda_{\rm b}$. 

At the nominal LHC design luminosity of $10^{34}$~\cms, multiple $\it pp$ collisions within the same bunch crossing (so-called pile-up) would significantly complicate the $b$ decay-vertex reconstruction and flavour tagging, and increase the combinatorial background.  For this reason the detector was designed to operate at a reduced luminosity. The luminosity at LHCb is locally controlled by transverse displacement of the beams (so-called ``levelling") to yield  $\lum = 4\times10^{32}\cms$ (approximately a factor of two above the original LHCb design value) at which the average event pile-up per visible crossing is $\sim$2. Running at relatively low luminosity has the additional advantage of reducing detector occupancy in the tracking systems and limiting radiation damage effects.

The dominant \bb-production mechanism at the LHC is through gluon-gluon fusion in which the momenta of the incoming partons are strongly asymmetric in the $\it pp$ centre-of-mass frame. As a consequence, the \bb\ pair is boosted along the direction of the higher momentum gluon, and both $b$ hadrons are produced in the same forward (or backward) direction in the $\it pp$ centre-of-mass frame. The detector is therefore designed as a single-arm forward spectrometer covering the pseudorapidity range $2<\eta<5$, which ensures a high geometric efficiency for detecting all the decay particles from one $b$ hadron together with decay products from the accompanying $\overline b$ hadron to be used as a flavour tag. The key detector features are a versatile trigger scheme efficient for both leptonic and hadronic final states, which is able to cope with a variety of modes with small branching fractions; excellent vertex and proper time resolution; precise particle identification, specifically for hadron ($\pi /K$) separation; precise invariant mass reconstruction to reject background efficiently. A full description of the detector characteristics can be found in Ref.\cite{ref:det}.

In this review, I report on a few recent precision measurements  of $CP$-violating observables in decays of beauty and charm-flavoured hadrons, while the study of rare decays at LHCb is covered by a separate contribution \cite{ref:Johannes}.

\section{Measurement of the \BS\ mixing phase}
The $CP$-violating phase $\phi_s$ between  \BBar\ mixing and the $b\rightarrow c\overline{c}s$ decay amplitude of the \BS\ meson is determined
with a  flavour tagged, angular analysis of the decay \BJPSIPHI, with $\Jpsi \to\mu^+\mu^-$ and $\phi \to K^+K^-$. This phase originates from the interference between the amplitudes for a \BS\ (\BbarS) to decay directly into the final state  $\Jpsi \phi$ or to first mix into \BbarS\ (\BS) and then decay.  In the SM, $\phi_s$ is predicted to be  $\simeq-2\beta_s$, where $2\beta_{\rm s}=2\rm{arg}(-V_{ts}{V^*}_{tb}/V_{cs}{V^*}_{cb})=(3.6\pm0.2)\,10^{-2}$~rad~\cite{ref:lenz}. However, NP could significantly modify this prediction if new particles contribute to the \BBar\ box diagram. 

The CDF and D0 collaborations~\cite{ref:CDFphis, ref:D0phis} have reported  measurements of the \BS\ mixing phase based on approximately 11,000 \BJPSIPHI\ candidates from an integrated luminosity of $9.6~\rm{fb}^{-1}$ (i.e. the full CDF Run II dataset) and 6,500 \BJPSIPHI\ candidates  from $8~\rm{fb}^{-1}$, respectively. Both results are compatible with the SM expectation at approximately one standard deviation in the ($\phi_s, \Delta\Gamma_s$) plane.  

LHCb has presented results  based on a sample that contains approximately 21,200 \BJPSIPHI\ candidates from an integrated luminosity of $1.0\rm\,{fb}^{-1}$~\cite{ref:LHCbphis}.
Besides having a very large and clean signal yield,  LHCb also benefits from an excellent proper time resolution to resolve fast \BS\ oscillations, which is measured to be $\sim$45~fs, compared to a \BS\ oscillation period of $\sim$350~fs. The other key experimental ingredient is flavour tagging, which is performed by reconstructing the charge of the $b$-hadron  accompanying the $B$ meson under study. 
The analysis uses an opposite-side flavour tagger based on four different signatures, namely high $\PT$ muons, electrons and kaons, and the net charge of an inclusively reconstructed secondary vertex, with a combined effective tagging power of $\sim$2.4\%.  The decay \BJPSIPHI\ is a pseudoscalar to vector-vector transition. Total angular momentum conservation implies $l=0,1,2$ and therefore the $\Jpsi \phi$ final state is a mixture of  $CP$-even ($l=0,2$) and $CP$-odd ($l=1$)  eigenstates, which can be disentangled on a statistical basis. This is accomplished by performing an unbinned maximum likelihood fit to the candidate invariant mass, decay time, initial \BS\ flavour  and the decay angles in the so-called transversity frame~\cite{ref:trans}.   The fit yields the following result for the three main observables, namely $\phi_s$, the decay width, $\Gamma_s$,  and the decay width difference between the light and heavy \BS\ mass eigenstates, $\Delta\Gamma_s$
\begin{align*}
\phi_s         &=-0.001&\pm\,&0.101    (\text{stat})&\pm\,& 0.027   (\text {syst})&\text {rad}, \\
\Gamma_s &=0.6580&\pm\,&0.0054  (\text{stat})&\pm\,& 0.0066 (\text {syst})&\text{ps}^{-1},\\
\Delta\Gamma_s &=0.116&\pm\,&0.018 (\text{stat})&\pm\,& 0.006(\text {syst})&\text{ps}^{-1}.
\end{align*}
This is the world's most precise measurement of $\phi_s$ and the first direct observation for a non-zero value for $\Delta\Gamma_s$. These results are fully consistent with the SM, indicating that $CP$ violation in the \BS\ system is small. Actually, there exists a second mirror solution in the plane $\Delta\Gamma_s$ vs $\phi_s$, which arises from the fact that the time-dependent differential decay rates are invariant under the transformation $(\phi_s,\Delta\Gamma_s) \rightarrow (\pi-\phi_s,-\Delta\Gamma_s)$ (plus an appropriate transformation of the strong phases). LHCb has recently resolved this ambiguity~\cite{ref:ambig} by studying the dependence of the strong phase difference between  the  $S$-wave and $P$-wave amplitudes on the $K^+K^-$ mass  from \BJPSIKK\ decays in the region around the $\phi(1020)$ resonance. The solution with positive $\Delta\Gamma_s$  is favoured with a significance of 4.7~standard deviations, indicating that in the \BS\ system the lighter $CP$ mass eigenstate that is almost $CP$ even decays faster than the state that is almost $CP$ odd.

The mixing-induced phase $\phi_s$ is also measured in the decay $B_s\to\Jpsi\pi^+\pi^-$~\cite{ref:Bs2Jpsipipi}. The branching fraction for this decay is $\sim 25\%$ of $B_s\to\Jpsi\phi$, with $\phi \to K^+K^-$. However, this final state has been shown to be almost $CP$ pure with a $CP$-odd fraction larger than 0.977 at 95\% CL,  and there is no need for an angular analysis.
About 7,400 signal events are selected in $1.0\,\rm{fb}^{-1}$ of data, yielding the result $\phi_s=-0.019^{+0.173+0.004}_{-0.174-0.003}$~rad.
The two datasets are combined in a simultaneous fit, leading to the preliminary result $\phi_s=-0.002\pm0.083\pm0.027$~rad~\cite{ref:LHCbphis}, in excellent agreement with the SM. The precision for this result is completely dominated by the statistical  uncertainty and therefore significant improvements are expected with more data.

\begin{figure}[!t]
\centering
\includegraphics[width=8cm]{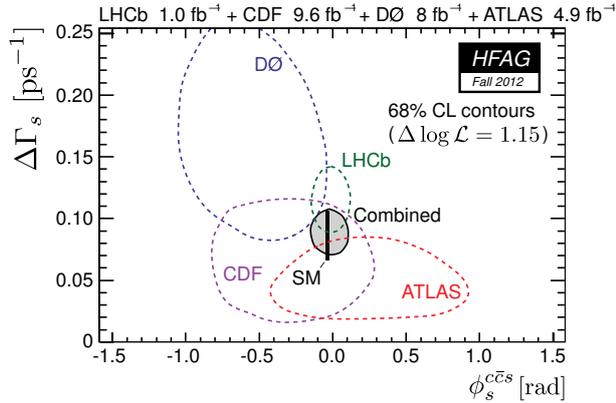}
\caption{\small 68\% confidence-level contours in the $(\phi_s,\Delta\Gamma_s)$ plane for the individual experimental measurements, their combined contour (solid line and shaded area), as well as the SM predictions (black rectangle). } 
\label{fig:phi_s}
\end{figure}

Figure~\ref{fig:phi_s} shows 68\%  confidence-level contours in the $(\phi_s,\Delta\Gamma_s)$ plane for the individual experimental measurements, their combined contour, as well as the SM predictions~\cite{ref:HFAG}. The combined result is consistent with these predictions at the 0.14~$\sigma$ level.

\section{Measurement of the weak phase $\gamma$ from tree-level decays}
\label{sec:gamma_from_trees}
The angle $\gamma$ is currently the least precisely known parameter of the CKM unitarity triangle. The direct determination of $\gamma$ via fits to the experimental data gives $(66\pm12)^{\circ}$~\cite{ref:CKM} or  $(76\pm10)^{\circ}$~\cite{ref:UTFIT}, depending on whether a frequentist or Bayesian treatment is used. In terms of the elements of the CKM matrix, this weak phase is defined as   $\gamma= {\rm arg}(-V_{\rm ud}{V^*}_{\rm ub}/V_{\rm cd}{V^*}_{\rm cb}) $. It is of particular interest as it can produce direct $CP$ violation in tree-level decays involving the interference between $b \rightarrow c\overline{u}s$ and $b \rightarrow u\overline{c}s$ transitions that are expected to be insensitive to NP contributions, thus providing a benchmark against which measurements sensitive to NP through loop processes can be compared.

One of the most powerful ways to measure the angle $\gamma$ is  through charged $B$ decays to open charm, $B^{\pm} \rightarrow Dh^{\pm}$, where  $D$ stands for a $D^0$ or a $\overline{D^0}$ and $h$ indicates either a pion or kaon.  Decays in which the hadron $h$ is a kaon carry greater sensitivity to $\gamma$.
The method is based on two key observations: 1. These decays  can produce neutral $D$ mesons of both flavours via colour-favoured or colour-suppressed decays  2. Neutral $D$ and $\overline{D}$  mesons can decay to a common final state, for example through Cabibbo-favoured or doubly Cabibbo-suppressed Feynman diagrams (ADS method~\cite{ref:ADS}) or through decays to $CP$ eigenstates such as $K^+K^-$ or $\pi^+\pi^-$ (GLW method~\cite{ref:GLW}). In the ADS case, the reversed suppression between $B$ and $D$ decays results in very similar amplitudes leading to a high sensitivity to $\gamma$. 
The relative phase between the two interfering amplitudes for  $B^{+} \rightarrow DK^{+}$  and $B^{+}\rightarrow \overline DK^{+}$ is the sum of the strong and weak interaction phases, while in the case of   $B^{-} \rightarrow DK^{-}$  and $B^{-}\rightarrow \overline DK^{-}$   it is the difference between the strong phase and $\gamma$. Therefore both phases can be extracted by measuring the two charge conjugate modes.  In addition, there is a dependence on the ratio between the magnitude of the suppressed amplitude and the favoured amplitude $r_B$.

\begin{figure}[!b]
\centering
\includegraphics[width=8cm]{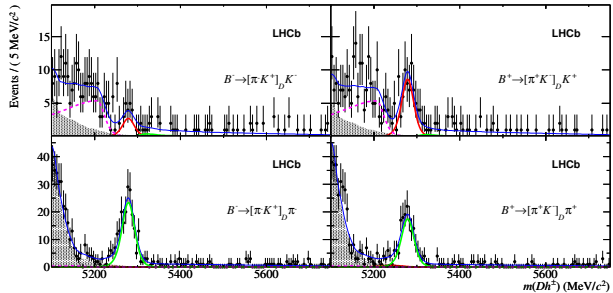}
\caption{\small Invariant mass spectra for  $B^{\pm} \rightarrow [\pi^{\pm}K^{\mp}]_Dh^{\pm}$  events; the left plots are $B^-$ candidates, $B^+$ are on the right. The dark (red) curve in the top plots represents the $B\rightarrow DK^{\pm}$ events, the light (green) curve in the bottom plots is $B\rightarrow D\pi^{\pm}$.} 
\label{fig:B2Dh}
\end{figure}

LHCb has performed an analysis of  the two-body $B^{\pm} \rightarrow Dh^{\pm}$ modes in which the considered $D$ meson final states are $\pi^+\pi^-$, $K^+K^-$, the Cabibbo-favoured $K^{\pm}\pi^{\mp}$ and the Cabibbo-suppressed $\pi^{\pm} K^{\mp}$~\cite{ref:BtoDh}.  For both GLW and  ADS methods the observables of interest are $CP$ asymmetries and partial widths. Exploiting $1.0\,\rm{fb}^{-1}$ of data, the first observation of the suppressed ADS mode $B^{\pm} \rightarrow [\pi^{\pm}K^{\mp}]_DK^{\pm}$ was performed. This is illustrated in the top plots of Fig~\ref{fig:B2Dh}. A large asymmetry is visible in the $B^{\pm} \rightarrow [\pi^{\pm}K^{\mp}]_DK^{\pm}$ mode (by comparing the signal yield for $B^+$ and $B^-$) while for  the  $B^{\pm} \rightarrow [\pi^{\pm}K^{\mp}]_D\pi^{\pm}$ ADS mode there is the hint of an (opposite) asymmetry. By combining all the various modes $CP$ violation in $B^{\pm} \rightarrow DK^{\pm}$ was observed with a $5.8\,\sigma$ significance. 

The three-body final state $D\rightarrow K^0_{\rm S}h^+h^-$,  with $h=(\pi,K$),  was also studied~\cite{ref:GGSZ} through a Dalitz plot analysis. The strategy relies on comparing the distributions of events in the $D\rightarrow K^0_s h^+h^-$ Dalitz plot for $B^+\rightarrow D K^+$ and  $B^-\rightarrow D K^-$ decays. Existing measurements of the CLEO-c experiment~\cite{ref:CLEO-c} were used to provide input on the $D$ decay strong-phase parameters. Based on approximately 800 $B^{\pm} \rightarrow DK^{\pm}$ decays with $D\rightarrow K^0_{\rm S}h^+h^-$ selected from $1.0\,\rm{fb}^{-1}$ of data, the following results were obtained for the weak phase $\gamma$ and the ratio $r_B$ between the suppressed and favoured $B$ decay amplitudes, $\gamma=(44^{+43}_{-38})^{\circ}$, with a second solution at $\gamma\rightarrow\gamma+180^{\circ}$, and $r_B=0.07\pm0.04$.

Other flavour specific final states, such as $D \rightarrow K\pi\pi\pi$ are exploited in a similar manner to the two-body case~\cite{ref:ADS4body}. However, for multi-body final states different intermediate resonances can contribute, diluting $CP$ violation effects. Based on $1.0\,\rm{fb}^{-1}$ of data, the suppressed ADS modes $B^{\pm}\rightarrow[\pi^{\pm}K^{\mp}\pi^+\pi^-]_DK^{\pm}$ and $B^{\pm}\rightarrow[\pi^{\pm}K^{\mp}\pi^+\pi^-]_D\pi^{\pm}$ were observed for the first time with significances of $5.1\,\sigma$ and $>10\,\sigma$, respectively.

A combination of  the $B^{\pm}\rightarrow DK^{\pm}$ results from Refs.~\cite{ref:BtoDh, ref:GGSZ, ref:ADS4body} was performed to derive an unambiguous best-fit value in $[0,180]^{\circ}$ of $\gamma=(71.1^{+16.6}_{-15.7})^{\circ}$~\cite{ref:gamma_comb},  suggesting very good prospects for the result based on the full data-set. Additional $\gamma$-sensitive measurements will also be included in the future. With the data currently available on tape, LHCb should be able to reduce the error quoted above by at least a factor of two.

\section{$CP$ violation in charmless $B$ decays}
Charmless $B$ decays represent an interesting family of channels for which a precise measurement of the charge or time-dependent $CP$ asymmetries can play an important role in the search for NP. In particular, NP may show up as virtual contributions of new particles in loop diagrams. A comparison of results from decays dominated by tree-level diagrams with those that start at loop level can probe the validity of the SM. In particular, it was pointed out by Fleischer~\cite{ref:fleischer} that $U$-spin related decays (obtained by interchanging $d$ and $s$ quarks) of the type $B_{s,d}\rightarrow h^+ h'^-$, with $h,h'=\pi,K$, offer interesting strategies for the measurement of the angle $\gamma$: in presence of NP in the penguin loops, such a determination could differ appreciably from that derived from  $B$ decays governed by pure tree amplitudes discussed in Sect.~\ref{sec:gamma_from_trees}.

\begin{figure}[!b]
\centering
\includegraphics[width=3.8cm]{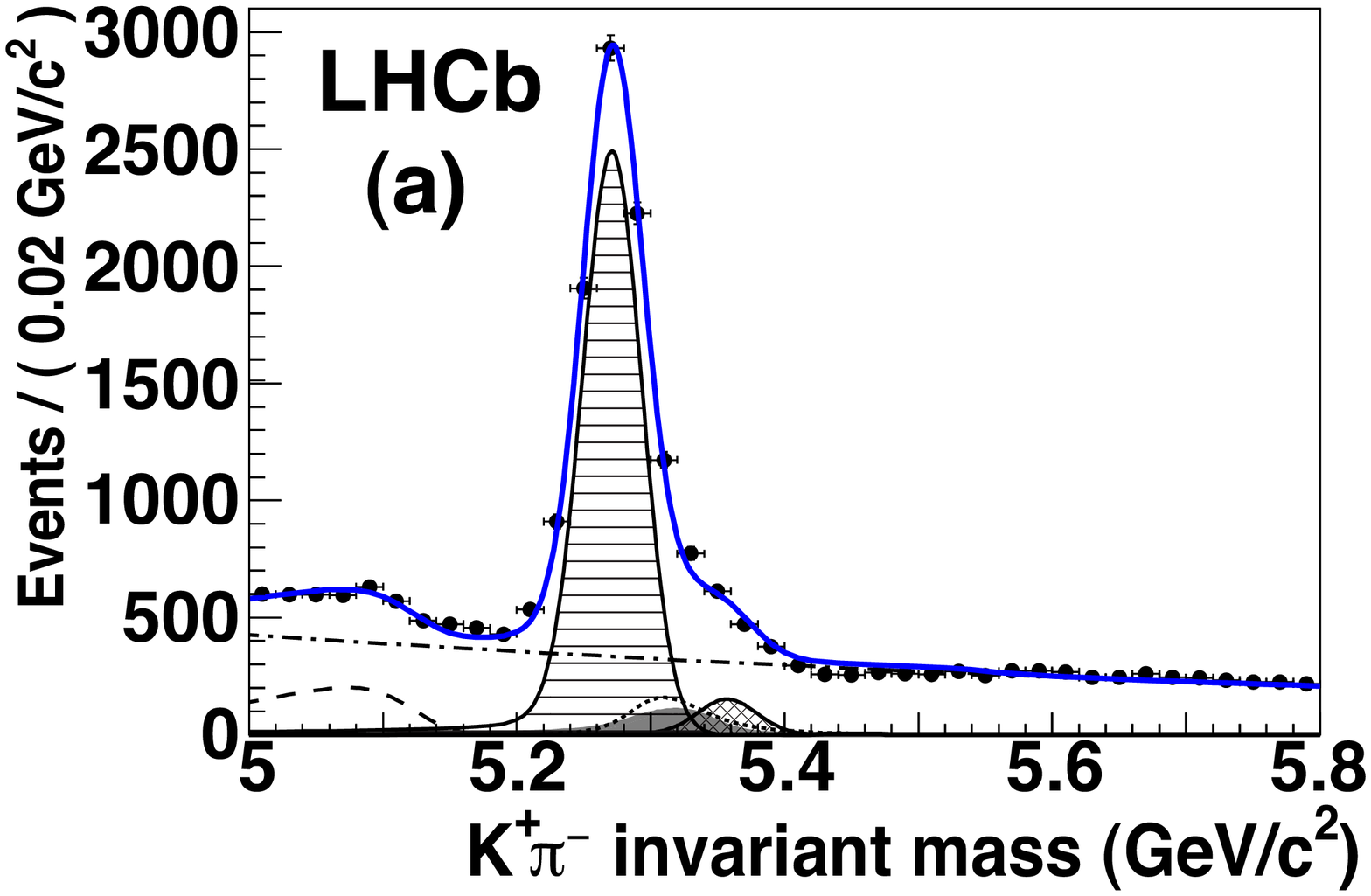}
\includegraphics[width=3.8cm]{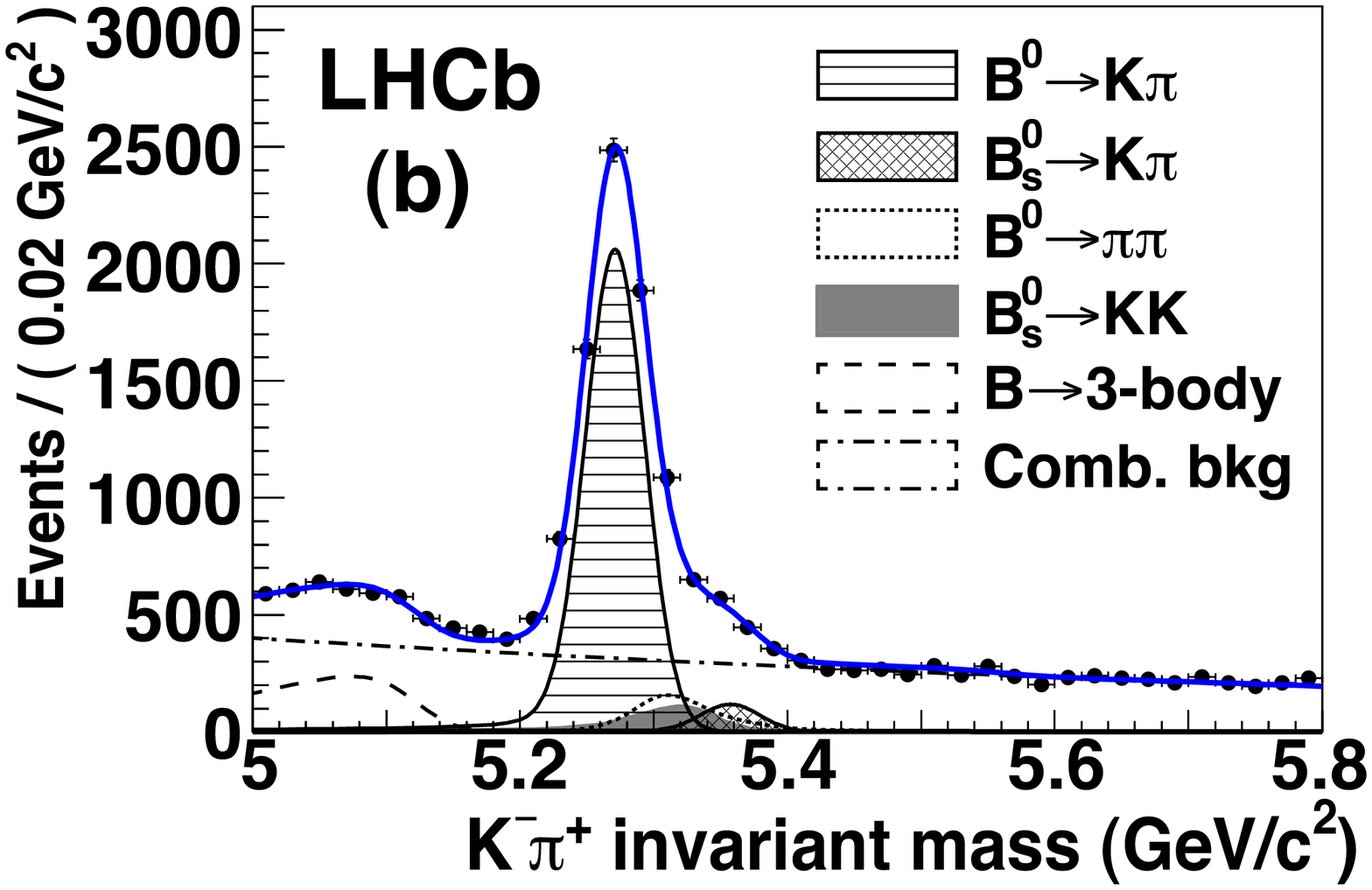}
\includegraphics[width=3.8cm]{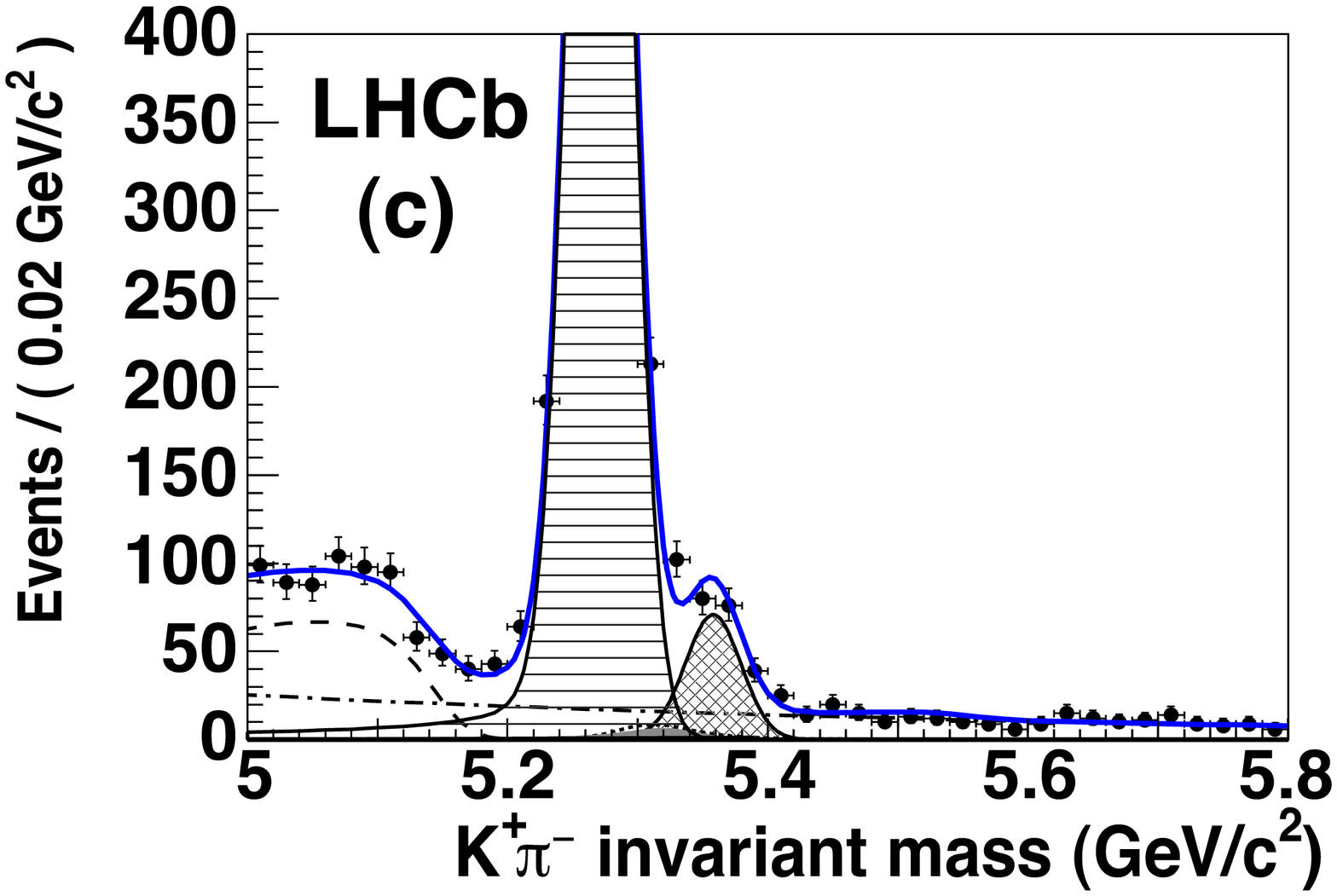}
\includegraphics[width=3.8cm]{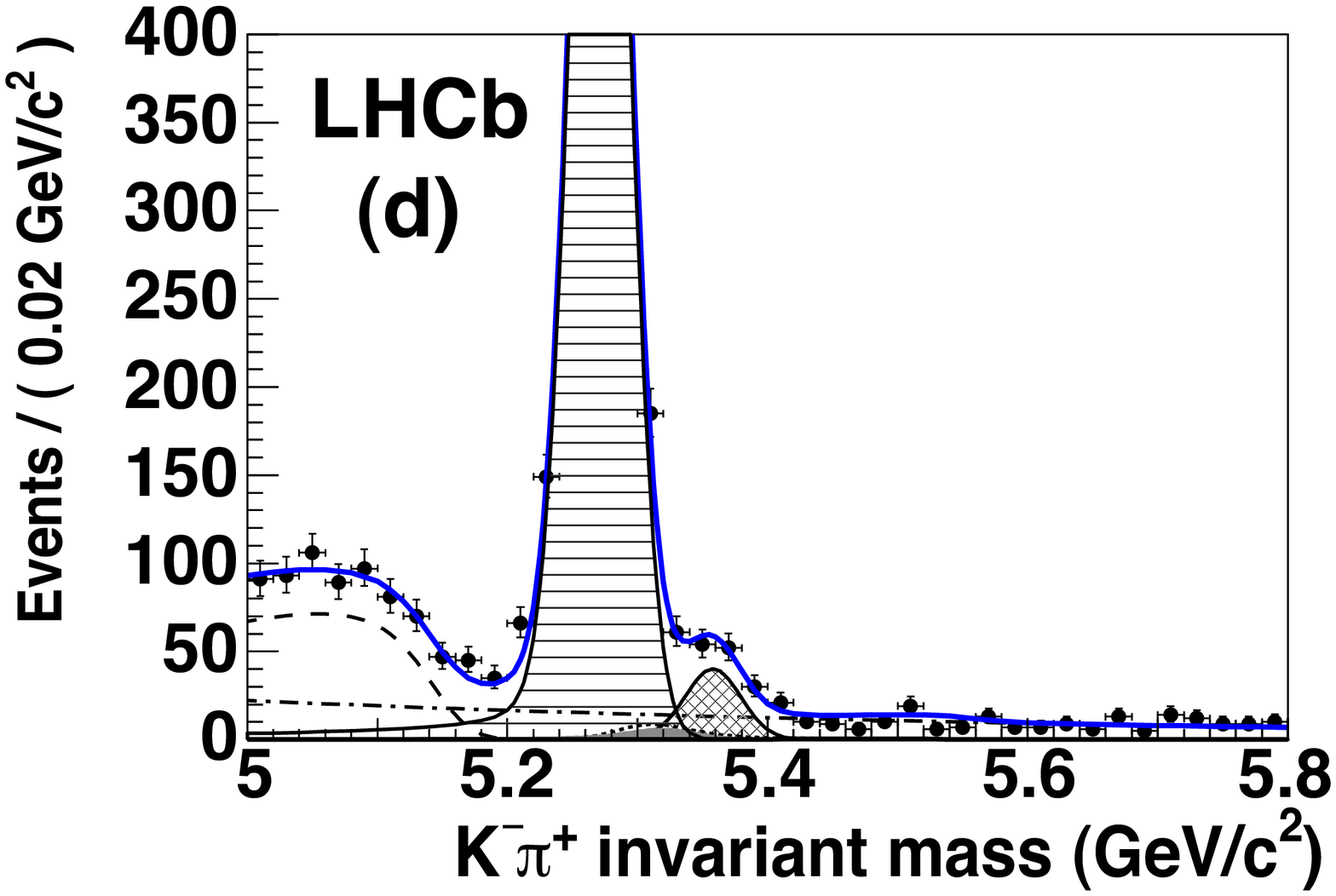}
\caption{\small Invariant $K\pi$ mass spectra for  $B_{s,d}\rightarrow K\pi$ events; the selection cuts are optimised for the best sensitivity to $A_{CP}(B_d\rightarrow K\pi$) (plots a and b) and $A_{CP}(B_s\rightarrow K\pi$) (c and d). Plots a and c (b and d) show the $K^+\pi^-$ ($K^-\pi^+$) invariant mass distribution. The main components of the fit model are also shown.} 
\label{fig:mass-plots}
\end{figure}

LHCb has analysed very large samples of $B_{s,d}\rightarrow h^+ h'^-$ decays,  separating the final state pions and kaons by using the particle identification  provided by the RICH detectors. Figure~\ref{fig:mass-plots} shows the invariant $K\pi$ mass spectra for $B_{s,d}\rightarrow K\pi$ events based on an integrated luminosity of $0.35\,\rm{fb}^{-1}$~\cite{ref:B2hh_int}. The selection cuts are optimised for the best sensitivity to $A_{CP}(B_d\rightarrow K\pi$) (plots a and b) and $A_{CP}(B_s\rightarrow K\pi$) (plots c and d), where the $CP$ asymmetry in the $B$ decay rate to the final state $f=K\pi$ is defined as $A_{CP}=\frac{(\Gamma(B\rightarrow f) -\Gamma(\overline B\rightarrow \overline f)  }{(\Gamma(B\rightarrow f) +\Gamma(\overline B\rightarrow \overline f) }$. LHCb reported~\cite{ref:B2hh_int} the most precise measurement for $A_{CP}(B_d\rightarrow K\pi)$ available to date, with a significance exceeding 6~$\sigma$, as well as the first evidence, at the 3.3~$\sigma$ level, for $CP$ violation in the decay of the $B_s$ mesons. The effect of the $CP$ asymmetry is visible in Fig.~\ref{fig:mass-plots} by comparing the yields  for the $B_d\rightarrow K\pi$ in (a) and (b) and those for the $B_s\rightarrow K\pi$ in (c) and (d).

The $CP$ asymmetry results need to be corrected for the effect of a possible $B$ production asymmetry, which is studied by reconstructing a sample of $B_d\rightarrow J/\psi K^*$ decays, given that $CP$ violation in $b\rightarrow c \overline{c} s$ transitions is expected to be small. Effects related to the instrumental detection efficiencies are evaluated by using data sets with opposite magnet polarities and reconstructing large samples of tagged $D^{*\pm}\rightarrow D^0(K^-\pi^+) \pi^{\pm}$ and $D^{*\pm}\rightarrow D^0(K^-K^+) \pi^{\pm}$ decays, as well as  untagged $D^0\rightarrow K^-\pi^+$ decays. These corrections are found to be small.

LHCb has also performed measurements of time-dependent $CP$ violation in charmless two-body $B$ decays by studying the processes $B_d\rightarrow \pi^+\pi^-$ and $B_s\rightarrow K^+K^-$~\cite{ref:B2hh_time}. The analysis is based on a luminosity of $0.69\,\rm{fb}^{-1}$. Direct and mixing-induced $CP$ asymmetries are measured in each channel using a tagged, time-dependent analysis. This analysis is performed for the first time at a hadron collider, and the $B_s\rightarrow K^+K^-$ decay is studied for the first time ever. The preliminary results for the  $B_d\rightarrow \pi^+\pi^-$ channel are in agreement with the world average from the $B$ factories. 
However, more data need to be analysed to be able  to extract a measurement of $\gamma$ from these  decays.

\section{$CP$ violation in charm}
The charm sector is an interesting place to probe for the presence of NP  because $CP$ violation is expected to be small in the SM. In particular, in singly Cabibbo suppressed decays, such as $D\rightarrow \pi^+\pi^-$ or $D\rightarrow K^+K^-$, NP could manifest itself through the interference between tree-level and penguin diagrams. LHCb has collected very large samples of charm: one in every ten LHC interaction results in the production of a charm hadron, of which 1-2~kHz are written to storage and are available for offline analysis. In 2012, LHCb has collected approximately $5\times 10^3$ tagged $D^{*\pm}\rightarrow (D^0\rightarrow K^+K^-)\pi^\pm$ and $3\times 10^5$ untagged $D\rightarrow K^-\pi^+$ decays per pb$^{-1}$ of integrated luminosity and it has the world's largest sample of two and three-body $D_{(s)}$ decays on tape.

The difference in time-integrated asymmetries between $D\rightarrow \pi^+\pi^-$ and $D\rightarrow K^+K^-$ was measured using $0.6\,\rm{fb}^{-1}$ of data collected in 2011~\cite{ref:delta_A_CP}. The flavour of the charm meson is determined by requiring a $D^{*\pm}\rightarrow (D^0\rightarrow h^+h^-)\,\pi_s^\pm$ decay, with $h=\pi$ or $K$, in which the sign of the slow pion $\pi_s$ tags the initial  $D^0$ or $\overline {D^0}$. By taking the difference of the measured time-integrated asymmetries for $\pi^+\pi^-$ and $K^+K^-$, effects related to the $D^*$ production asymmetry and to the detection asymmetry of the slow pion and  $D$ meson in the final state cancel to first order, so that one can derive the difference of the $CP$ asymmetries, $\Delta A_{CP}$. Second-order effects are minimised by performing the analysis in bins of the relevant kinematic variables. A nice additional advantage of taking this difference is that in the U-spin limit $ A_{CP}(KK)=-A_{CP}(\pi\pi)$ for any direct $CP$ violation~\cite{ref:implications}, so that the effect is amplified. The final result is
\begin{equation}
\Delta A_{CP}=[-0.82\pm0.21\,(\rm stat)\pm0.11\,(\rm syst)]\%.
\end{equation}
This result (subsequently confirmed by CDF~\cite{ref:CDF_delta_A_CP}) has generated a great deal of theoretical interest, as it is the first evidence for $CP$ violation in the charm sector, with a significance of $3.5~\sigma$. 

The difference in time-integrated asymmetries can be written to first order as
\begin{equation}
\Delta A_{CP}=[a^{dir}_{CP}(K^+K^-)-a^{dir}_{CP}(\pi^+\pi^-)]+\frac{\Delta \langle t \rangle}{\tau}a^{ind}_{CP}\,,
\end{equation}
where $a^{dir}_{CP}$ is the asymmetry arising from direct $CP$ violation in the decay, $\langle t \rangle$ the average decay time of the $D^0$ in the reconstructed sample, and $a^{ind}_{CP}$ the asymmetry from $CP$ violation in the mixing. In presence of a different time acceptance for the $\pi^+\pi^-$ and $K^+K^-$ final states, a contribution from indirect $CP$ violation remains.

\begin{figure}[!t]
\centering
\includegraphics[width=8cm]{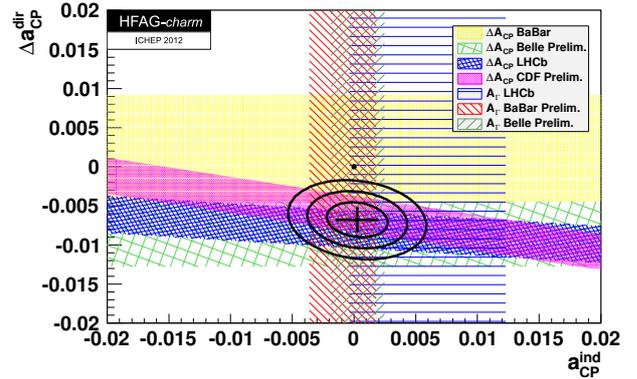}
\caption{\small HFAG combination of $\Delta A_{CP}$ and $A_\Gamma$ measurements~\cite{ref:HFAG_Delta_ACP}, where the bands represent $\pm1\sigma$ intervals. The point of no $CP$ violation (0,0) is displayed as a black dot; the ellipses show the two-dimensional 68\% CL, 95\% CL and 99.7\% CL with the best fit value as a cross.} 
\label{fig:HFAG_Delta_ACP}
\end{figure}

Figure~\ref{fig:HFAG_Delta_ACP} shows the HFAG world-average combination~\cite{ref:HFAG_Delta_ACP} in the plane ($a^{ind}_{CP},\,a^{dir}_{CP}$). The combined data is consistent with no $CP$ violation at 0.002\% CL.

LHCb is currently completing the analysis of the full 2011 data sample and pursuing alternative strategies to verify the effect. In particular, an analysis is being finalised in which the flavour of the $D^0$ meson is tagged using the charge of the muon in semileptonic $B$ decays; furthermore  $CP$ violation is searched for in charged $D$ decays where there is no possibility of indirect $CP$ violation and a positive signal would indicate unambiguously the presence of direct $CP$ violation.
\section{Conclusion}
This contribution has reviewed measurements of $CP$ violation in charm and beauty decays performed by LHCb using up to $1.0\,\rm{fb}^{-1}$ of data collected in 2011. In the $B^0_s$ system LHCb has obtained the world's most precise measurement of the mixing phase $\phi_s$  and the first direct observation for a non-zero value for  $\Delta\Gamma_s$. Important milestones have been reached in the measurement of the weak phase $\gamma$ both from decays at tree level and from those where new physics could contribute through loops. The large charm production cross-section at the LHC has allowed for a dramatic improvement in sensitivity to $CP$ violating effects. During the 2012 run at $\sqrt{s}=8$~TeV, another $2.1\rm{fb}^{-1}$ of data were collected, which will allow LHCb to increase the precision on the results presented here and pursue new analyses  to understand precisely the nature of $CP$ violation in charm and beauty and hopefully find signs of new physics.


\section*{Acknowledgement}
I would like to thank the organisers of Capri 2012 for their kind invitation to such a nice and fruitful meeting on a truly splendid island, my LHCb colleagues for providing the material discussed here and, in particular, Tim Gershon and Vava Gligorov for their careful reading of this article.

\nocite{*}
\bibliographystyle{elsarticle-num}
\bibliography{martin}



\end{document}